\documentclass[conference,10pt]{IEEEtran}
\ifCLASSINFOpdf
  \usepackage[pdftex,final]{graphicx}
  \usepackage{epstopdf}
\else
\fi
\usepackage{bm}
\usepackage{dsfont}
\usepackage{amssymb,amsmath}
\usepackage{mathtools}

\usepackage{subcaption}

\usepackage[hyphens]{url}

\newcommand{\newprocess}[1]{\ensuremath{#1 = (#1 _t)_{0 \leqslant t \leqslant T}}}
\newcommand{\newprocessinf}[1]{\ensuremath{#1 = (#1 _t)_{t \geqslant 0}}}
\newcommand{\running}[3]{#1=#2,\,\ldots,#3}

\newcommand{\thetavector}{\bm{\theta}}

\newcommand{\thetaestimateml}{\widehat{\thetavector}_{\mathrm{ML}}}

\newcommand{\Mprocess}{M^H}

\newcommand{\psiprocessn}{\bm{\psi}^H}
\newcommand{\Rfunc}{\bm{R}_H}

\newcommand{\fbm}{B^H}

\newcommand{\timeset}{t \geqslant 0}

\newcommand{\firsttime}[1]{\ensuremath{\inf \{\timeset: #1\}}}

\newcommand{\bbbr}{{\rm I\!R}} 

\begin{document}
%
\title{Detecting Performance Degradation\\
of~Software-Intensive Systems in~the~Presence\\
of~Trends and Long-Range Dependence}

\author{
 \IEEEauthorblockN{Alexey Artemov}
 \IEEEauthorblockA{Lomonosov Moscow State University\\
 Complex Systems Modeling Laboratory,\\
 27-1 Lomonosovsky Ave., Moscow 119991, Russia}
 \IEEEauthorblockA{Yandex Data Factory,\\
 16 Leo Tolstoy St., Moscow 119021, Russia,\\
 Email: artemov@physics.msu.ru}\\
 \and
 \IEEEauthorblockN{Evgeny Burnaev}
 \IEEEauthorblockA{Skolkovo Institute of Science and Technology,\\
 3 Skolkovo Innovation Center,
 Moscow, 143026, Russia}
  \IEEEauthorblockA{Institute for Information Transmission Problems,\\
 19 Bolshoy Karetny Lane, Moscow 127994, Russia,\\
 Email: e.burnaev@skoltech.ru}
 }


%


\maketitle

\begin{abstract}
As contemporary software-intensive systems reach
increasingly large scale, it is imperative that
failure detection schemes be developed to help prevent
costly system downtimes. A promising direction
towards the construction of such schemes is
the exploitation of easily available measurements 
of system performance characteristics such as average number
of processed requests and queue size per unit of time.
In this work, we investigate a holistic methodology
for detection of abrupt changes in time series data
in the presence of quasi-seasonal trends and long-range dependence
with a focus on failure detection in computer systems.
We propose a trend estimation method enjoying
optimality properties in the presence of long-range dependent noise
to estimate what is considered ``normal'' system behaviour.
To detect change-points and anomalies,
we develop an approach based on the ensembles of ``weak'' detectors.
We demonstrate the performance of the proposed change-point
detection scheme using an artificial dataset,
the publicly available Abilene dataset as well
as the proprietary geoinformation system dataset.
\end{abstract}


%
\IEEEpeerreviewmaketitle

\section{Introduction}
\label{introduction}

The last decade has witnessed the emergence of a novel
type of high-tech systems: the software-intensive
systems~\cite{iso42010}. The
latter\footnote{defined in ISO/IEC/IEEE 42010:2011 as
systems where ``software contributes essential
infuences to the design, construction, deployment, and evolution
of the system as a whole''} include digital
communication systems, internet systems (including devices,
data transfer networks and services), call centers, 
automated power grids, intellectual transport systems,
electronic trading platforms and many others.
The obvious requirement for such systems
is the effective, reliable and uninterrupted operation.
However, recent studies of large-scale software-intensive systems
indicate quite the opposite state of affairs: due to their sheer
scale\footnote{Expressed in ``number of lines of code; number
of people employing the system for different
purposes; amount of data stored, accessed, manipulated,
and refined; number of connections and interdependencies
among software components; and number of hardware
elements''~\cite{northrop2006ultra}.}
``software and hardware failures will be the norm
rather than the exception''~\cite{Yigitbasi2010}.
According to the research, the dominant cause
of costly and dangerous system failures are the software failures
which makes software ``the most problematic element
of large-scale systems''~\cite{northrop2006ultra}.

Among the efforts undertaken in order to improve
system reliability a major role is played by failure
detection which aims to identify failures
based on the analysis of data collected during
the system operation. Examples of such data include
the average number of processed requests and the queue
size per time unit, the volume of transferred traf{}fic,
the number of rejected queries, etc.
During both unexpected events (such as network equipment
failures and network attacks) and scheduled occasions (e.\,g.
data center maintenance and system software upgrades)
the data experience abrupt deviations from the target state.
The goal then is to detect sudden changes (referred to as anomalies
or disorders) in the f{}low of the observed data. 
The detection is to be performed online;
within the online (sequential) setting,
as long as the behavior of the observations is consistent
with the target state, one is content to let the process
continue. If the state changes, then one is interested
in detecting the change as rapidly as possible.
Problems concerned with constructing efficient procedures
for detecting changes in observed stochastic processes are known
in the literature as change-point detection problems~\cite{Polunchenko2012}.

\begin{figure*}
\begin{center}
\centerline{
\includegraphics[
    width=0.70\textwidth,
    trim=3cm 0 4cm 0.5cm,
    clip=true
]{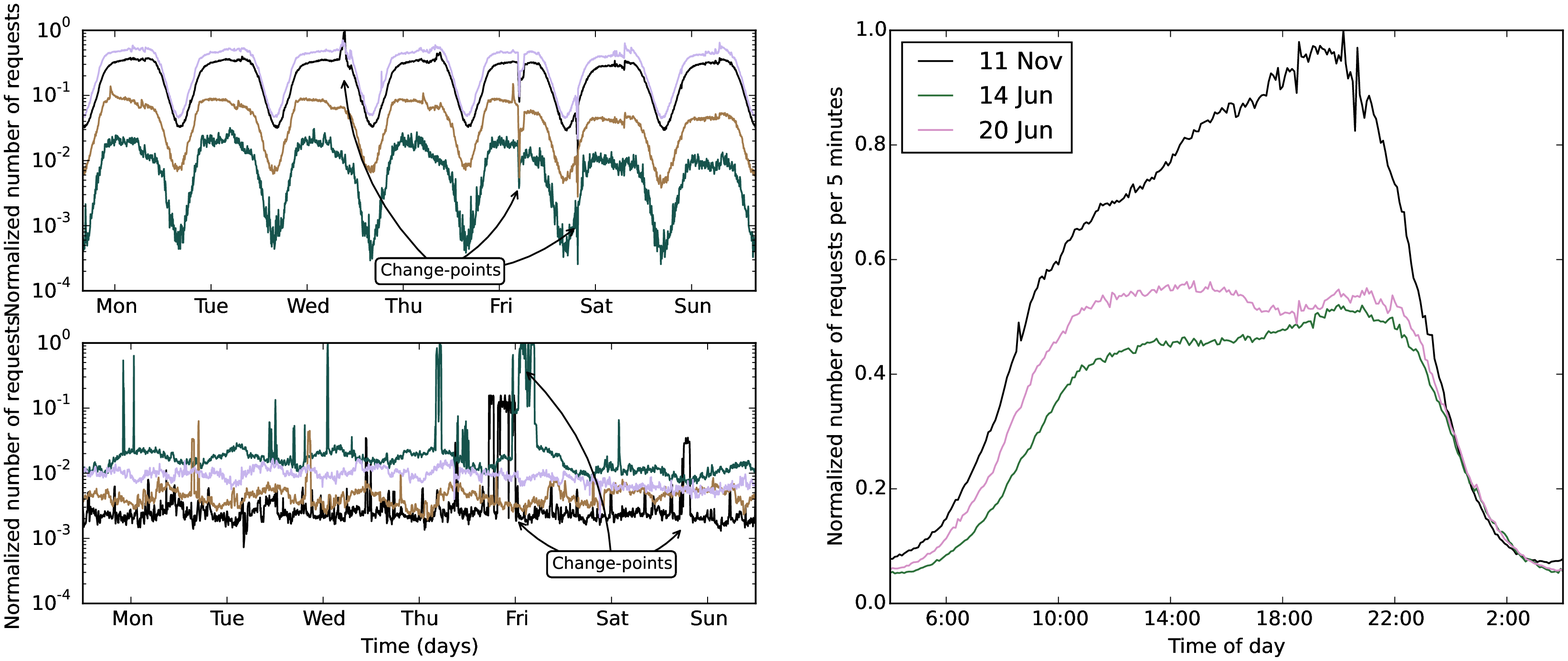}
}
\caption{Top-left: weekly load profile of a geoinfomation system at Yandex
along with several change-points. Right: daily load
of a system at Yandex aggregated over
consecutive 5-minute intervals for three different days in 2014:
Saturday, 15th June, Friday, 20th June, and
Tuesday, 11th November. Note the change of the
load profile from weekday to weekend and throughout the year.
Bottom-left: weekly load shape of the traffic in the
Abilene network (1008 measurements),
for the period of 14--21 June, 2004, along with several
change-points.}
\label{fig:data_example}
\end{center}
\vskip -0.2in
\end{figure*}

In the present work,
we investigate the change-point detection problem for localization
and diagnosis of anomalies in large-scale software-intensive systems
in the presence of quasi-periodic trends and long-range dependence.
The key step in the change-point detection approach is the specification
of what is ``normal'' and ``abnormal'' state.
This problem represents a challenge due to a number of reasons.
First, the systems we consider here experience anthropogenic ``nearly
periodic'' load variations that are difficult to model
due to a complex load shape and its random variations over time.
An example of quasi-periodic time series we investigate in this work
is shown in Fig.~\ref{fig:data_example}; they reflect weekly and daily
load profiles for several internet services.

The second essential property of data flows in large-scale
computer systems is that long-term correlations are present
in these quantities, i.\,e., they are statistically
self-similar~\cite{Leland1994}. As self-similarity
(also referred to as \textit{long-range dependence} or \textit{LRD})
has significant impact on queueing performance
and represents the dominant cause of load ``bursts'',
the model should be able to ef{}ficiently capture it.

A natural approach to change-point detection involves
utilization of statistical detection procedures
such as the CUSUM procedure~\cite{Page1954}, 
the control charts procedure~\cite{Shewhart1931}, etc., as
they possess certain ef{}ficiency properties.
In turn, for these procedures to be implemented,
the change-point model (the model of the ``normal'' and the ``abnormal''
signals) must be specified.
The latter often cannot be specified accurately; as a consequence,
even theoretically optimal procedures suf{}fer significant
degradation in change-point performance.

Finally, a considerable dif{}ficulty is caused by the large scale
of contemporary software-intensive systems.
For instance, volume of the dataset measured at
Yandex\footnote{Yandex is one of the largest internet companies in Europe,
operating Russia's most popular search engine and its most
visited website, see \url{http://company.yandex.com}.}
reaches hundreds of thousands of characteristics,
while other authors report software systems consisting of
up to tens of thousand nodes~\cite{Yigitbasi2010}.
As the cost of manual model selection for each individual
observed signal might be unacceptable, one should consider
an automatic approach to model learning.

In this paper, we present an optimal method for signal estimation
and an efficient procedure for change-point and anomaly detection
in the presence of quasi-periodic trends and long-range dependence.
We use our theoretical results regarding the structure
of the optimal filter to construct a practical
trend estimation algorithm. Using the estimate, we develop
the change-point detection algorithm based on the ensemble
of ``weak'' detectors to improve change-point detection
performance when the standard assumptions regarding
the change-point model are violated.

We briefly describe existing change-point detection approaches
as well as some of the conventional filtering techniques
in Sect.~\ref{related_work}. In Sect.~\ref{model_spec}, we specify our
time series model and propose the model estimation algorithm.
In Sect.~\ref{changepoint_detection}, we consider the particular
change-point detection problem for our model and develop 
the ensemble-based change-point detection method.
Sect.~\ref{applications} presents the evaluation results
for a simulated and two real-world datasets: a publicly available Abilene
network dataset and a proprietary Yandex dataset.

\section{Related Work}
\label{related_work}

A vast body of research covers the problem of failure detection
in computer systems, and efficient detection algorithms have been developed
for anomaly detection in computer networks, data stream networks, etc,
see, e.\,g.,~\cite{Pham2014,Casas2010} and references therein.
In these applications, the change-point detection problem
is investigated for the case of stationary random series,
which is a well-studied setting (See~\cite{Polunchenko2012} for a bird's eye review).

Process stationarity assumption is rather restrictive
for practice since in many applications the observed
process is non-stationary. While no specific assumptions
about the structure of the observed process are made,
purely data-driven approaches such as principal
component analysis (PCA) and its modifications are often
taken under consideration~\cite{Casas2010}. PCA
and the subspace methods classify the observed data into ``normal''
and ``abnormal'' subspaces and have proven themselves
efficient in anomaly detection
applications~\cite{Pham2014,Lakhina2004,Casas2010}.

Change-point detection approaches mentioned above are
difficult to apply directly to our problem. On the one hand,
the observed data in our system are non-stationary;
on the other hand, these data are characterized by trends 
and LRD noise that make PCA and the subspace methods ineffective.

A body of research covers a vast number of trend modeling
and estimation approaches, such as
multiple exponential smoothing~\cite{winters1960forecasting},
autoregressive models~\cite{findley1998new},
decomposition methods~\cite{hodrick1997postwar},
parametric and nonparametric
regression~\cite{artemov2015nonparametric}.
Neither of the approaches incorporates an explicit model
of LRD; consequently, efficient trend estimation
in the presence of LRD cannot be achieved. 
On the contrary, our trend extraction approach relies
on an explicit model of LRD signal
and yields theoretically efficient estimates.

\section{Trend Estimation in the Presence\\
of Long-Range Dependence}
\label{model_spec}


\subsection{LRD and the Fractional Brownian Motion}
\label{fbm_theory}

Long-range dependence is a phenomenon shared by many
natural and technical systems. It relates to the rate of decay
of statistical dependence of points with increasing time interval.
In relation to software-intensive systems, LRD
may be qualified as the presence of ``burstiness'' across an extremely
wide range of time scales~\cite{Leland1994}. During the last
decades, the fractional Brownian motion has been established
as the standard model for LRD signals.

The fractional Brownian motion (fBm) was introduced by Kolmogorov
in connection with his works on the theory
of turbulence~\cite{kolmogorov1940wiener} and later was
constructively defined by Mandelbrot~\cite{Mandelbrot1968}.
In what follows, we adopt the notation from~\cite{artemov2015optimal}.
A standard fBm~$\newprocess{\fbm}$
with Hurst exponent $H\in\left(0,1\right)$ on $\left[0,T\right]$
is a Gaussian process with continuous trajectories, $\fbm_0 = 0,$ $\mathbf{E}\fbm_t = 0,$ $\mathbf{E}\fbm_s \fbm_t =
  \frac{1}{2}\left(t^{2H} + s^{2H} - \left|t-s\right|^{2H}\right).$ When $H=\frac{1}{2}$, the process $\fbm$ is a standard Brownian
motion but in the case $H\neq\frac{1}{2}$ the process~$\fbm$
is not a semimartingale.
In many applications, process $\fbm$ is used for modeling of time series
with very chaotic movements (the case $H < 1/2$) and
with a relatively smooth behavior (the case $H \geqslant 1/2$).

\subsection{The Specification of the Theoretical Filter}
\label{filter_theory}

Let the observed continuous-time process~$\newprocess{X}$ satisfy the relation
\begin{equation}
X_{t}=\sum\limits _{i=0}^{n}\theta_{i}\varphi_i(t) + \sigma B_{t}^{H},
\label{eq:observation_explicit_form}
\end{equation}
where $\{\varphi_i(t)\}_{i=0}^{n}$ is a dictionary
of differentiable functions on $[0, T]$,
$\newprocess{\fbm}$ is the standard fBm on $[0, T]$ with a known Hurst index $H$,
and the variance $\sigma > 0$ is assumed to be known.
The unknown parameters $\{\theta_i\}_{i=0}^n$ must be estimated
using the observations $\left\{X_{s},0\leqslant s\leqslant t\right\}$
available up to time $t$.

In~\cite{artemov2015optimal}, theoretical results
regarding the structure of the optimal filter in~\eqref{eq:observation_explicit_form}
for the general dictionary of functions $\{\varphi_i(t)\}_{i=0}^{n}$
were obtained for the case of (a) the maximum likelihood estimate
and (b) the Bayesian estimate.
For the purpose of the current work, we use the maximum likelihood (ML) filter
to estimate a smooth trend against the LRD noise. We assume that:
\begin{itemize}
\item the dictionary consists
of power functions: $\varphi_i(t) = t^i, i=0, \ldots, 3$,
allowing to estimate the polynomial trend~$f(t)$;
\item the value of the Hurst exponent~$H$ is known (in practice,~$H$
can be estimated from the observations using
such approaches as introduced in~\cite{kirichenko2011comparative,hardstone2012detrended});
\item the value of the variance $\sigma$ is known
(in fact, the filter from~\cite{artemov2015optimal}
does not depend on the variance, see below).
\end{itemize}
The ML estimate $\thetaestimateml$ for the drift
parameter~$\thetavector=(\theta_0, \ldots, \theta_3)$ is given by
\begin{equation}
\label{eq:ml_estimate_generic}
\thetaestimateml = \Rfunc^{-1}(t) \psiprocessn_t,
\end{equation}
where~$\Rfunc(t) = (\Rfunc(t))_{ij}$ and
$\psiprocessn_t = ((\psiprocessn_t)_0, \ldots, (\psiprocessn_t)_3)$
are defined by $(\Rfunc(t))_{ij} = \alpha_H(i,j) t^{i+j-2H}$ and
$(\psiprocessn_t)_i = \beta_{H}(i) \int\limits _{0}^{t}s^{i-1}d\Mprocess_s,$
where $\lambda_{H} = 2H\frac{\Gamma(3-2H) \Gamma(1/2 + H)} {\Gamma(3/2-H)},$ 
$\alpha_H(i,j) = \lambda^{-1}_{H}\beta_{H}(i)\beta_{H}(j)\frac{2-2H}{i+j-2H},$
$\beta_{H}(i) = i \tfrac{2-2H+i-1}{2-2H}\tfrac{\Gamma\left(3-2H\right)}{\Gamma\left(3-2H+i-1\right)} \tfrac{\Gamma\left(3/2-H+i-1\right)}{\Gamma\left(3/2-H\right)},$ $ \running{i,j}{0}{n},$
and $\newprocess{\Mprocess}$ is a martingale defined by
$\Mprocess_t \equiv \kappa_{H}^{-1} \int\limits _{0}^{t} s^{1/2-H}(t-s)^{1/2-H}d X_s,$
$\kappa_{H} = 2H\Gamma(3/2 - H)\Gamma(1/2 + H) \enspace.$

\subsection{The Trend Estimation Algorithm with LRD Correction}
\label{ml_estimation_algorithm}

The algorithm assumes the observations 
are taken according to the model
\begin{equation}
\label{eq:general_observations_model}
X_t = f(t) + \eta^H(t), \qquad t \geqslant 0,
\end{equation}
where the trend $f(t)$ is some smooth function observed in 
the LRD noise~$\eta^H(t)$. Taking advantage of the smoothness
of the trend $f(t)$, we approximate it using
some finite-order polynomial $\sum_{i = 0}^n \theta_i (t - t_0)^i$ in the neighbourhood
of any $t_0 > 0$. We model~$\eta^H(t)$ using
the fractional Gaussian noise (fGn) $Z_t^H$ with some (unknown but nonrandom)
variance~$\sigma(t)$ and Hurst exponent~$H$: $\eta^H(t) = \sigma(t) Z_t^H$.
Given the noisy observations $\{(X_k, t_k)\}_{k = 1}^{\ell}$,
the goal is to estimate the expected
value $f(t) = \mathbf{E} X_t$ for any $t \geqslant 0$.
The following algorithm provides a solution to this problem.

\begin{enumerate}
\item Consider an interval $[a,b]$ and select
observations~window~$W(a, b) = \{(X_k, t_k): a \leqslant t_k \leqslant b\}$.

\item Compute the estimate $\widehat{f}_{[a,b]}(t)$ of the trend $f(t)$
for $a \leqslant t \leqslant b$:

\begin{enumerate}
\item Assuming a cubic polynomial model for the observations
\label{fbm_algorithm:theta_estimate}
\begin{equation}
\label{eq:locpoly_observations_model}
X_k = \sum \limits_{i = 0} ^3 \theta_i (t_k - t_0)^i + \sigma Z_k^H,
\end{equation}
where $(X_k, t_k) \in W(a, b)$, $t_0 = (a + b) / 2$,
$\sigma$~is assumed to be constant, and $H = \frac{1}{2}$,
 estimate the value
of~$\thetavector = (\theta_0, \ldots, \theta_3)$
using the maximum likelihood estimate~$\thetaestimateml$
described in Sect.~\ref{filter_theory}.

\item 
\label{fbm_algorithm:trend_estimate}
Compute the trend estimate on $[a,b]$ using the relation
$\widehat{f}_{[a,b]}(t) = \sum _{i = 0} ^3 (\thetaestimateml)_i (t - t_0)^i$
for each $t \in [a, b]$.

\item
\label{fbm_algorithm:var_estimate}
Compute the variance estimate $\widehat{\sigma}$
as the sample variance of residuals
$\{X_k - \widehat{f}_{[a,b]}(t_k) \mid t_k \in [a, b] \}$.

\item Compute the estimate of the Hurst exponent $\widehat{H}$
using an approach from~\cite{hardstone2012detrended}
and the standardized residuals
$\{(X_k - \widehat{f}_{[a,b]}(t_k)) / \widehat{\sigma} \mid t_k \in [a, b] \}$.

\item Using the Hurst exponent estimate $\widehat{H}$,
compute corrected trend and variance estimates in~a)--c).
\end{enumerate}

\item We use the sliding window $[a,b]=[a, a + \Delta]$ with sufficiently large $\Delta$
and obtain $n_{[a,b]}(t) = \left\vert{A(t)}\right\vert$ local
corrected estimates~$\widehat{f}_{[a,b]}(t)$ for each $t \geqslant 0$,
where $A(t) = \{(a, b) \mid t \in [a, b]\}$.
To obtain the final estimate $\widehat{f}(t)$ 
we average the corrected estimates using the relation
$\widehat{f}(t) = \frac{1}{n_{[a,b]}(t)} \sum\limits_{(a,b)\in A(t)} \widehat{f}_{[a,b]}(t) \enspace.$
\end{enumerate}

\begin{figure}[t!]
\centering
\begin{subfigure}[t]{\columnwidth}
\centering
\includegraphics[
    width=0.6\columnwidth,
]{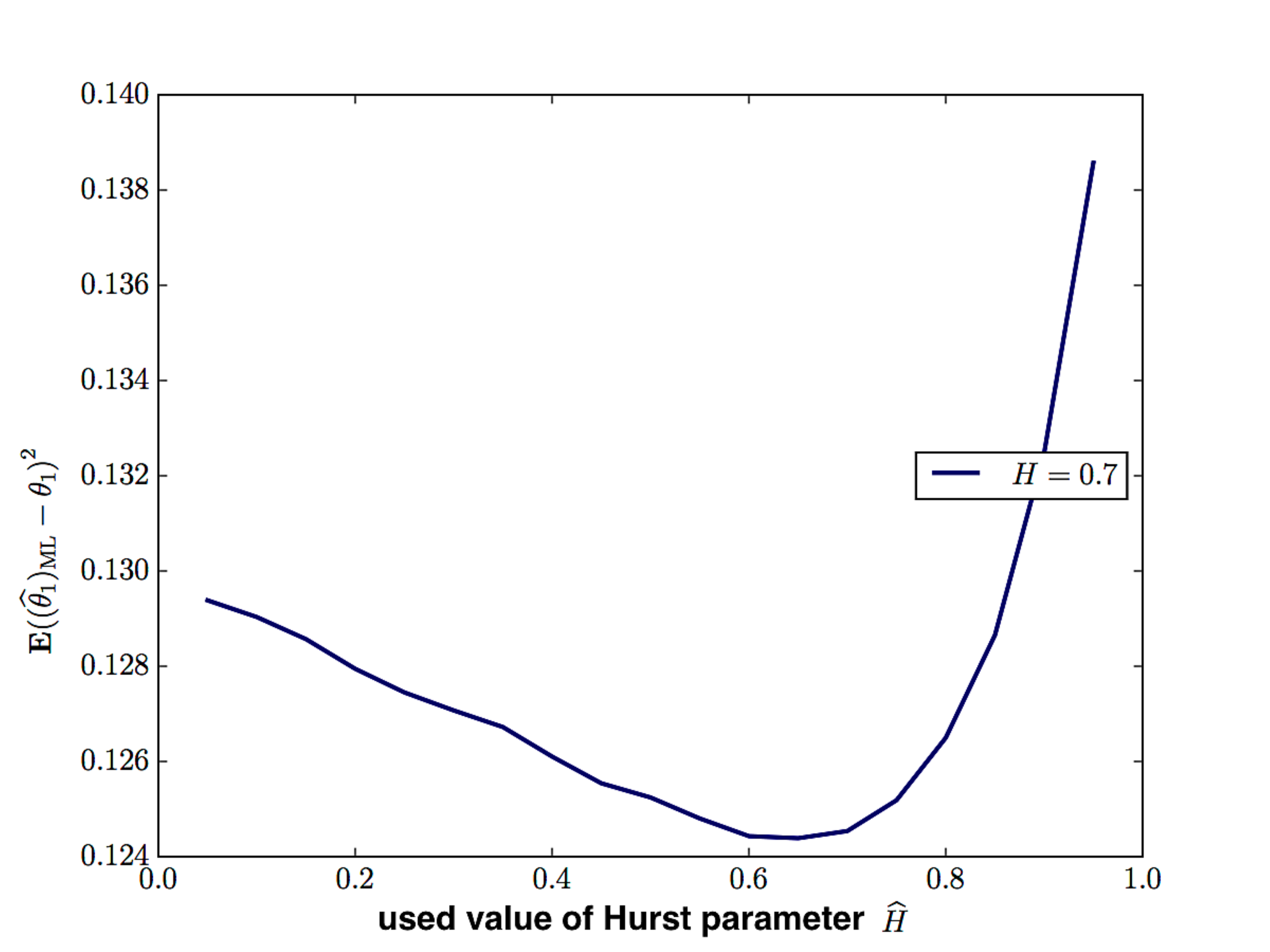}
\caption{Influence of the Hurst exponent value used in the algorithm
in Sect.~\ref{ml_estimation_algorithm} on the trend estimation performance.
The results are obtained using $10^6$ replications of Monte-Carlo
and rescaled to $[0,1]$ for better viewing.}
\label{fig:mc_trend_filter_vs_hurst}
\end{subfigure}
\\
\begin{subfigure}[t]{\columnwidth}
\centering
\includegraphics[
    width=0.6\columnwidth,
]{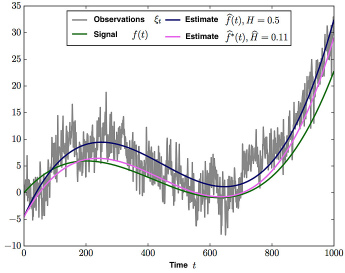}
\caption{The effect of the correction in~\ref{fbm_algorithm:theta_estimate}--\ref{fbm_algorithm:var_estimate}
 on the trend estimation accuracy. The estimate $\widehat{f}(t)$ was obtained using $\widehat{H} = 0.5$, while the corrected estimate $\widehat{f}^*(t)$ was obtained using $\widehat{H} = 0.11$ (true $H=0.1$).}
\label{fig:mc_correction_effect}
\end{subfigure}
\caption{Correction employed in the algorithm in Sect.~\ref{ml_estimation_algorithm}
and its effect on the trend extraction performance.}
\end{figure}

The two-step procedure for computing the estimate~$\widehat{f}(t)$
is necessary since in practice the Hurst exponent is unknown
but important constant strongly influencing an estimation performance,
see Fig.~\ref{fig:mc_trend_filter_vs_hurst}. By applying the correction in
the algorithm steps~\ref{fbm_algorithm:theta_estimate}--\ref{fbm_algorithm:var_estimate}
we achieve better trend estimation accuracy compared
to a generic approach with $H = \frac{1}{2}$, see~Fig.~\ref{fig:mc_correction_effect}.

\section{Change-point Detection in the Presence\\
of Trends and Long-Range Dependence}
\label{changepoint_detection}


\subsection{The Change-point Model}

We consider the following change-point model for the noise~$\eta^H(t)$
in~\eqref{eq:locpoly_observations_model}:
\begin{equation}
\label{eq:short_term_change_process}
\eta^H(t) = \mu \mathds{1}_{[\theta, \theta + \Delta t]}(t) + \sigma Z^H_t,
\qquad t \geqslant 0,
\end{equation}
where $\theta$ is an unknown time of a change,
$\mu$ is an unknown change magnitude, 
$\sigma$ is an unknown (non-random) variance,
and $Z^H_t$ is the fGn.
The characteristic duration $\Delta t$ of the considered change 
is short; hence the change represents a local deviation
in the values of the observed series, see Fig.~\ref{fig:trajectory_real}.

To detect the change, we introduce a residual process $\newprocessinf{R}$
\begin{equation}
\label{eq:residual_change_process}
R_t = \sigma^{-1} (X_t - \widehat{X}_t),
\qquad t \geqslant 0,
\end{equation}
where $X_t$ is the signal with a known variance $\sigma$
observed in~\eqref{eq:locpoly_observations_model}
and $\widehat{X}_t$ is an estimate of $X_t$ obtained
via filtering algorithm described in Sect.~\ref{ml_estimation_algorithm}.
In absence of a change, $R_t$ is an approximately zero-mean process
with unit variance, however, in presence of a change,
neither of these properties holds.
Note that $R_t$ is a natural estimate for $Z^H_t$
and $\sigma R_t$ is a natural estimate
for the noise component~$\eta^H(t)$.
We use the process $R_t$ in Sect.~\ref{change_point_detection} to detect the change.

\subsection{The Ensemble-based Change-point Detection Procedure}
\label{change_point_detection}

The standard assumptions regarding
the change-point model state that pre- and post-change
distributions are Gaussian i.i.d. with different (yet known)
parameters~\cite{Shewhart1931,Page1954,Polunchenko2012}.
These assumptions are heavily violated in our case
due to (a) the approximation error introduced
by substitution of the real trend $f(t)$ with a locally
cubic trend, (b) the estimation error introduced
by the estimation algorithm in Sect.~\ref{ml_estimation_algorithm},
(c) the unknown change signature, and (d) the modeling errors
due to interpreting noise in the real signal as the fBm.
Moreover, the absence of accurate detection procedures
for LRD signals makes the change-point detection performance low
when ``classical'' change-point detection methods are used.

Let $\Pi_1, \ldots, \Pi_n$ denote $n$ change-point detection procedures, 
such as the cumulative sum (CUSUM) procedure~\cite{Page1954}
based on the process~$\newprocessinf{T}$:
\begin{equation}
\label{eq:cusum_definition}
T_t  = 
  \max(0,T_{t - 1} + \zeta_t),
  \quad T_0 = 0, \quad \timeset,
\end{equation}
where $\zeta_t = \log (f_0(X_t) / f_{\infty}(X_t))$
is the log-likelihood ratio, and $f_{\infty}(\cdot)$ and $f_0(\cdot)$ 
are one-dimentional pre- and post-change distributions, respectively.
Each procedure $\Pi_k$ prescribes to stop observations at time~$\tau_k$
which is the first hitting time of some process~$\newprocessinf{S^k}$
to a level~$h_k > 0$: $\tau_k = \firsttime{S^k_t \geqslant h_k}$.
We further consider a set of \textit{signals}
$\big\{\newprocessinf{s^k}\big\}_{k = 1}^{n}$
defined by $s^k_t = S^k_t / h_k, \timeset$.
We call the procedure~$\mathrm{A}$ an \textit{ensemble} if its stopping time~$\tau_{\mathrm{A}}$
is defined as the first hitting time of some process~$\newprocessinf{a}$ to a specified
level~$h_{\mathrm{A}} > 0$: $\tau_{\mathrm{A}} = \firsttime{a_t \geqslant h_{\mathrm{A}}}$, where 
\begin{equation}
\label{eq:general_ensemble}
a_t = \psi(\bm{\lambda}; \mathbf{S}^1_t, \ldots, \mathbf{S}^n_t),
\end{equation}
$\bm{\lambda} \in \bbbr^d$ ($d\geqslant n$) and
$\mathbf{S}^k_t = \{s^k_s, 0 \leqslant s \leqslant t\}$
is the history of the signal~$\newprocessinf{s^k}$ up to the
time $t$, $k = 1, \ldots, n$.
Each ensemble is completely defined by the choice of
the ``aggregation function'' $\psi(\cdot)$. 
In this work, we consider a \textit{logistic regression-based}
classifier for which the aggregation function could be written as
\begin{equation}
\label{eq:logistic_weighted_p_ensemble}
a_t = \psi_{\textsc{Log}-p}(\bm{\lambda}; \mathbf{S}^1_t, \ldots, \mathbf{S}^n_t) = 
  \sigma\Big(\sum\limits _{j = 0} ^{p} \sum\limits _{k = 1} ^{n} \lambda_{kj} s^k_{t - j} - \lambda_0\Big),
\end{equation}
where $\sigma(x) = 1 / (1 + e^{-x})$ is the logistic function.
The value $a_t$ can be interpreted as a posterior probability
of a change-point given the observations history
$\mathbf{X}_t = \{X_s, 0 \leqslant s \leqslant t\}$ up to the moment $t$.
Note that for this ensemble the threshold $h_{\mathrm{A}}$ must be
chosen to belong to the interval~$(0, 1)$~\cite{artemov2015ensembles}.

\subsection{Learning Ensemble Parameters}

Ensemble parameters $\bm{\lambda} \in \bbbr^d$
can be \textit{learned} to optimize a certain performance measure.
Let~$\mathcal{X}^{\ell} = \{(X^i, Y^i)\}_{i=1}^{\ell}$ be the labeled data where
each point~$(X^i, Y^i) \in \mathcal{X}^{\ell}$ is a pair,
its first component~$\newprocess{X^i}$
being a sample path of the observations, and its label~$\newprocess{Y^i}$ being
an ``abnormal'' state indicator: $Y^i_t = \mathds{1}_{\mathcal{T}^{i}_{0}}(t)$. 
Let $T_{\infty}^i$ and $T_0^i$ be the durations of ``normal'' and
``abnormal'' states $\mathcal{T}^{i}_{\infty}$ and $\mathcal{T}^{i}_{0}$
for each point~$(X^i, Y^i), i=1, \ldots, \ell$, respectively.
We formulate the problem of learning the parameters~$\bm{\lambda} \in \bbbr^d$
of an ensemble as an optimization
problem~$\mathbf{F} (\mathrm{A}) \to \inf \limits_{\bm{\lambda} \in \bbbr^d}$
for the Average Relative Error Rate measure 
\begin{multline}
\label{eq:ara_measures}
\mathbf{F} (\mathrm{A}) = 
c_{\infty}
\mathbf{E}_{\infty}
    \Bigg[ 
      \frac {
            \int \mathds{1}_{\{a_t \geqslant h_{\mathrm{A}}\}}(t) \mathds{1}_{\mathcal{T}_{\infty}}(t) dt
            }
            {
            \int \mathds{1}_{\mathcal{T}_{\infty}}(t) dt
            }
    \Bigg] +  \\
c_{0}
  \mathbf{E}_{0}
    \Bigg[ 
      \frac {
            \int \mathds{1}_{\{a_t < h_{\mathrm{A}}\}}(t) \mathds{1}_{\mathcal{T}_{0}}(t) dt
            }
            {
            \int \mathds{1}_{\mathcal{T}_{0}}(t) dt
            }
    \Bigg],
\end{multline}
where~$c_{\infty}$ and~$c_{0}$ are
the costs of false alarm and false silence, respectively.
As $\mathbf{F} (\mathrm{A})$ is a non-differentiable function and 
cannot be optimized using standard approaches, we introduce its
empirical approximation~$\widehat{\mathbf{F}}_{\mathrm{D}}(\mathrm{A})$
defined by
\begin{multline}
\label{eq:empirical_ara_upper_bound}
\widehat{\mathbf{F}}_{\mathrm{D}}(\mathrm{A}) =
    \frac{1}{\ell} \sum\limits_{i=1} ^{\ell} 
    \Bigg\{
      \frac {c_{\infty}} {T^i_{\infty}}
      \sum\limits_{t \in \mathcal{T}^i_{\infty}} \sigma(a_t - h_{\mathrm{A}})
    + \\
      \frac {c_{0}} {T^i_{0}}
      \sum\limits_{t \in \mathcal{T}^i_{0}} \sigma(h_{\mathrm{A}} - a_t),
    \Bigg\}
\end{multline}
where $\sigma(x) = 1 / (1 + e^{-x})$ is the logistic function. Note now
that the function~$\widehat{\mathbf{F}}_{\mathrm{D}}(\mathrm{A})$
is differentiable w.\,r.\,t. the ensemble parameters~~$\bm{\lambda} \in \bbbr^d$
and can therefore be optimized using standard methods.

\section{Performance Evaluation}
\label{applications}

\subsection{Evaluation Datasets}
\label{sec:evaluation_datasets}

We study the performance of filtering and change-point
detection algorithms on two artificial datasets
\textsc{Artificial-Easy} and \textsc{Artificial-Hard} and on two
real-world datasets: the publicly available Abilene network dataset
and on the proprietary Yandex dataset.

Artificial datasets consist of one-week samples
of artificial data $\{(X_k, t_k)\}_{k = 1}^{K}$, $K = 2016,$
measured at consecutive 5-minute intervals 
according to the model $X_k = f(t_k) + \eta^H(t_k)$,
where $f(t_k) = A \sin(2\pi t_k / T)$ with $A = 1.5, T = 288$,
and $\eta^H(t)$ is the LRD noise process.
To model the change-point in the artificial data,
for each replication of the sample we generate the LRD noise $\eta^H(t)$
according to the model in~\eqref{eq:short_term_change_process}
with $\sigma = 1$, a random change-point time: $\theta \sim U(T, 6T)$, 
a random change-point duration: $\Delta t \sim U(5, 100)$,
and $\newprocessinf{Z^H}$ formed as a discrete approximation of the fGn process
with $H = 0.95$. For \textsc{Artificial-Easy}, we set the change-point
magnitude $\mu = 5$, and for \textsc{Artificial-Hard}, change-point
magnitude is set to $\mu = 3$. Despite this seemingly large
magnitude, as we show below, the change-points we generated
are remarkably hard to detect, due to the presence of seasonal trends and LRD noise, see Fig.~\ref{fig:season_approx} (left).
We generated 1000 independent replications of the sample
for training the ensemble and another 1000 for testing.
We denote these dataset $\mathcal{X}^{\ell}_{\mathrm{TRAIN}}$ and
$\mathcal{X}^{\ell}_{\mathrm{TEST}}$, where $\ell = 1000$, respectively. 

The Abilene dataset\footnote{See \url{http://www.cs.utexas.edu/~yzhang/research/AbileneTM}.} describes network
load in the Abilene network in terms of the amount
of traffic transmitted between network
endpoints during consecutive 5-minute intervals. The data is available
for the period of March 1, 2004 to September 10, 2004, and consists
of 132 different time series describing traffic transmitted between
12 different network nodes located in 12 different locations across
the USA. An example of Abilene data is shown in Fig.~\ref{fig:data_example},
bottom-left, for 4 different pairs of endpoints for a particular
measurement period. The Abilene dataset is frequently used for evaluation
of anomaly detection methods due to its complex
structure and presence of both short-lived and long-lived
anomalies~\cite{Lakhina2004,Casas2010}.

The Yandex dataset consists of time series
describing the performance of a geoinformation system at Yandex.
Each time series is sampled at consecutive 5-minute intervals and
it represents the total number of requests processed by the system.
An example of Yandex time series is shown in Fig.~\ref{fig:data_example} (top-left) and in Fig.~\ref{fig:trajectory_real} (right) along with labels displaying the anomalies subject to detection.

\subsection{Evaluated Procedures}
\label{sec:evaluated_procedures}

We train the ensemble using five ``weak'' detectors:
the cumulative sum detector, the Shiryaev-Roberts detector,
the Shewhart detector, the changepoint detector, and
the posterior probability process detector (for details,
refer to~\cite{artemov2015ensembles}, Sect.~2).

We empirically compare the performance of our ensemble-based
procedure to that of several well-studied approaches, specifically,
threshold-based procedure, CUSUM procedure, and the subspace method.
The threshold-based procedure \textsc{EWMA-Threshold} uses
EWMA to estimate the mean~$\widehat{\mu}_t$ and variance~$\widehat{\sigma}^2_t$
of the time series~$X_t$, obtains the residuals
$R_t = (X_t - \widehat{\mu}_t) / \widehat{\sigma}_t$, and calculates the fraction
of the residual points within the time window $[t - \Delta, t]$
located above the threshold~$h$. The stopping time for raising
the alarm is defined as $\tau_{\mathrm{THR}} = \inf \{k \geqslant 1:
S_k \geqslant h_{\mathrm{THR}}\}$ where $S_k =
\sum_{i = k - \Delta}^k \mathds{1}_{ \{R_i \geqslant h\} } (i)$.
The threshold $h_{\mathrm{THR}}$, the per-point threshold $h$ and the window size~$\Delta$
are algorithm parameters; we only report results regarding
the calibrated values of these parameters which result in best performance
of the procedure. The \textsc{EWMA-CUSUM} procedure replaces the statistic $\newprocessinf{S}$
defined above with the CUSUM statistic~$T_t$ defined in~\eqref{eq:cusum_definition}.
The densities $f_{\infty}(\cdot)$ and $f_0(\cdot)$ are assumed to be normal
with unit variances and means $\mu_{\infty} = 0$ and $\mu_0 = \mu_{\infty} + \delta$,
respectively. The parameter $\delta$ is selected to obtain the best performance
on training set in terms of the area under the precision-recall curve.
The subspace method \textsc{PCA} is closely related to the singular spectrum
analysis (SSA) approach and subspace methods from the
literature~\cite{Lakhina2004,Casas2010,vautard1992singular}.
In the \textsc{PCA} procedure, a decomposition of the time series $\newprocessinf{X}$
is obtained using the SSA procedure, and the component $\mathbf{X}^{\mathrm{RES}}_t$
living in the residual subspace is considered. The statistic $\newprocessinf{P}$
of the procedure is the norm of the residual component:
$P_t = \Vert\mathbf{X}^{\mathrm{RES}}_t\Vert$.
We note that the subspace method benefits greatly from pretraining
on historic data. To exploit this advantage, we supplied the SSA procedure
with a week of historic data to obtain a better decomposition.
We call this procedure \textsc{PCA-Pretraining}. Note that no other
procedure receives any additional input when trained.

\subsection{Trend Approximation Accuracy}
\label{sec:trend_extraction_accuracy}

\begin{figure}[h!t!]
\begin{center}
\centerline{
\includegraphics[
    width=0.6\textwidth,
    trim=-2.4cm 0cm 0cm 0cm,
    clip=true
]{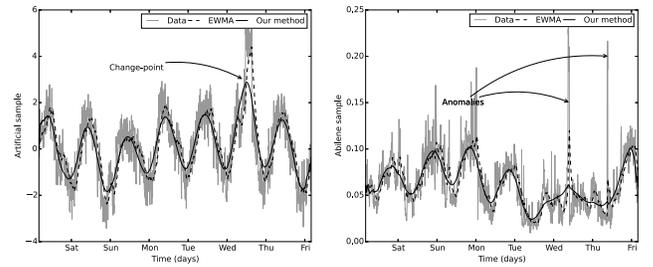}
}
\caption{Example data from the \textsc{Artificial-Easy} dataset
(left) and the Abilene dataset (right)
and trend extraction results obtained using EWMA and our approach.
Marked are the labeled anomalies.
}
\label{fig:season_approx}
\end{center}
\end{figure}

We first compare the trend extraction accuracy on the dataset \textsc{Artificial-Easy}.
We use the relative root mean squared forecast error $\mbox{RRMSE}(X_t, \widehat{X}_t) = \sqrt{\frac{1}{K} \sum_{t=1}^K (X_t - \widehat{X}_t)^2/X_t^2},$
to evaluate forecasting performance. Table~\ref{table:trend_approximation_rmse} presents
trend extraction accuracy on two tasks: trend approximation and one-point-ahead forecasting.
Trend approximation accuracy $\mbox{RRMSE}(f(t), \widehat{f}(t))$ measures how closely
the extracted trend follows the true trend $f(t)$. One-point-ahead forecasting
accuracy estimates how well an algorithm predicts incoming new data $X_t$
given the observed values $\{X_k, k < t\}$. Our study shows that our approach
produces significantly more accurate estimates than EWMA.
An example of trend approximation is presented in Fig.~\ref{fig:season_approx}
for the artificial dataset and for the Abilene dataset. We note that our
approach yields a smooth approximation and allows for more robust anomaly
isolation, while EWMA follows the data more closely.

\begin{table}[h]
\caption{Trend extraction accuracy for the artificial dataset
in terms of RRMSE (\%) for EWMA, PCA and our approach.}
\label{table:trend_approximation_rmse}
\vskip 0.15in
\begin{center}
\begin{small}
\begin{sc}
\begin{tabular}{lcc}
\hline
Method & \begin{tabular}[c]{@{}c@{}}Trend\\approximation\end{tabular} &
\begin{tabular}[c]{@{}c@{}}One-point-ahead\\forecasting\end{tabular} \\
\hline
EWMA  & 7.84 & 7.34 \\
PCA   & 8.96 & 5.65 \\
PCA-Pretraining   & 5.58 & 3.80 \\
Ours  & 5.72 & 3.06 \\
\hline
\end{tabular}
\end{sc}
\end{small}
\end{center}
\vskip -0.1in
\end{table}

\begin{figure}[t!h!]
\begin{subfigure}{0.52\textwidth}
\centering
  \includegraphics[width=\linewidth]{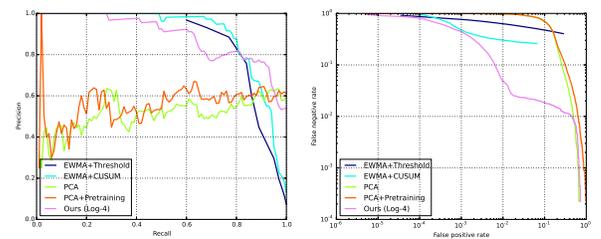}
  \caption{\textsc{Artificial-Easy} dataset}\label{fig:detection_artificial_easy}
\end{subfigure}\hfill
\begin{subfigure}{0.52\textwidth}
\centering
  \includegraphics[width=\linewidth]{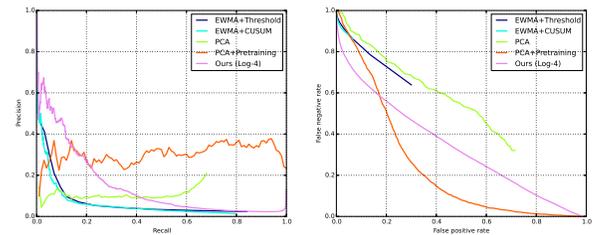}
  \caption{\textsc{Artificial-Hard} dataset}\label{fig:detection_artificial_hard}
\end{subfigure}
\caption{Empirical comparison of change detection performance for EWMA-based approaches, PCA-based approaches, and our approach. Left: Precision-Recall curves.
Right: Average Relative Error Rate curves.}
\end{figure}

\subsection{Change-point Detection Performance Measures}
\label{sec:change_point_detection_results}

To evaluate the change-point detection performance, we use two
performance measures. The first measure is the Precision-Recall Curve,
which is a standard performance measure in the area of machine learning.
The second measure is the Average Relative Error Rate curve
proposed in~\eqref{eq:ara_measures}--\eqref{eq:empirical_ara_upper_bound}. 
Before discussing the obtained results, 
we briefly explain how these performance measures are computed.
Suppose that a procedure $\Pi$ is defined by a statistic $\newprocessinf{S}$.
When computed on a test instance $(X^i, Y^i) \in \mathcal{X}^{\ell}_{\mathrm{TEST}}$,
procedure $\Pi$ generates a~trajectory $\{S^i_1, \ldots, S^i_l\}$
and for some specified threshold $h_{\mathrm{\Pi}} > 0$ produces 
$M_{\mathrm{\Pi}}$ segments $\big\{[t_{a_m}, t_{b_m}]\big\}_{m=1}^{M_{\mathrm{\Pi}}}$ such that
$\forall t \in [t_{a_m}, t_{b_m}] \quad S^i_t \geqslant h_{\mathrm{\Pi}}$.
We declare the detection $[t_{a_m}, t_{b_m}]$ true positive if it
intersects with the ``abnormal'' segment, i.\,e. if
$[\theta, \theta + \Delta t] \cap [t_{a_m}, t_{b_m}] \neq \emptyset$.
If, on the other hand, this intersection is empty (the statistic
signals outside the interval $[\theta, \theta + \Delta t]$),
then the detection is declared false positive. The Precision-Recall Curve
is plotted by varying the threshold $h_{\mathrm{\Pi}}$. The 
Average Relative Error Rate curve is a plot of Average False Positive Rate
$\frac{1}{\ell} \sum_{i=1} ^{\ell}
      \frac {c_{\infty}} {T^i_{\infty}}
      \sum_{t \in \mathcal{T}^i_{\infty}} \mathds{1}_{\{S^i_t \geqslant h_{\mathrm{\Pi}}\}}(t)$
versus Average False Negative Rate
$\frac{1}{\ell} \sum_{i=1} ^{\ell} 
      \frac {c_{0}} {T^i_{0}}
      \sum_{t \in \mathcal{T}^i_{0}} \mathds{1}_{\{S^i_t < h_{\mathrm{\Pi}}\}}(t)$.
Average Relative Error Rate can be thought of as a segmentation
rather than classification measure.

\subsection{Results}
\label{sec:experiments_results}

For the \textsc{Artificial-Easy} data, our approach is outperformed only
by the optimal CUSUM procedure by a little margin when measured in
terms of AUC, see Fig.~\ref{fig:detection_artificial_easy}, left.
On \textsc{Artificial-Hard}, our approach outperforms
all other methods in equal conditions. However, adding more
data to \textsc{PCA} to improve decomposition accuracy
makes it the best on this task, see~Fig.~\ref{fig:detection_artificial_hard}, left.
Our approach also yields the most accurate segmentations,
as can be seen on both Fig.~\ref{fig:detection_artificial_easy}, right,
and Fig.~\ref{fig:detection_artificial_hard}, left, meaning
both lower average false silence and lower average false alarm durations.

\begin{figure*}[t!h!]
\begin{center}
\centerline{
\includegraphics[
    width=0.7\textwidth,
    trim=3cm 0 4cm 0.5cm,
    clip=true
]{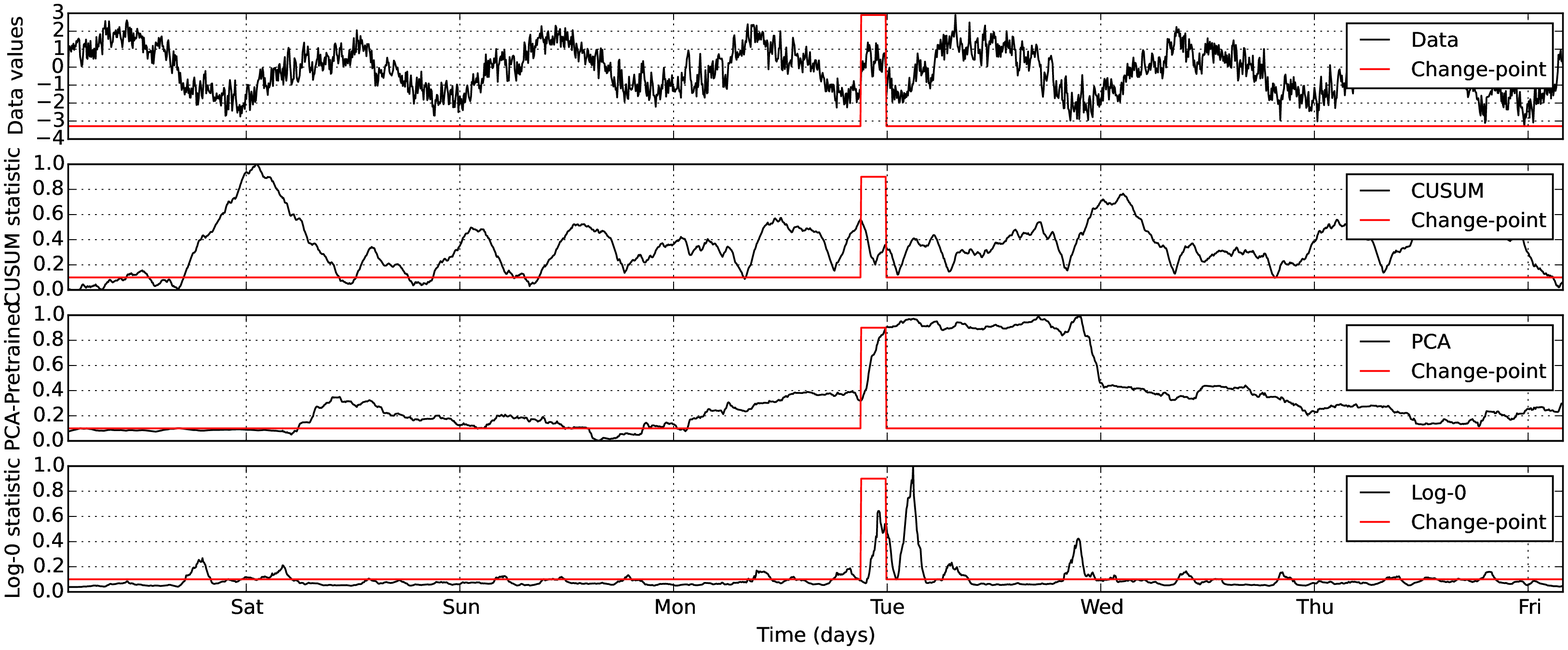}
}
\caption{Comparison of CUSUM, \textsc{PCA-Pretrained} and \textsc{Log-0} ensemble
trajectories for the \textsc{Artificial-Hard} dataset.
Top: a sample of artificial data with the change-point indicator.
Upper middle: sample path of the CUSUM statistic. Note no correlation between the two.
Lower middle: sample path of the \textsc{PCA-Pretrained} statistic.
Note the longer duration of the large values of the statistic.
Lower: sample path of the logistic regression-based
classifier statistic along with the change-point indicator.
Note the strong correlation between the two.
All statistics have been rescaled to $[0, 1]$ for better viewing experience.}
\label{fig:trajectory_artificial}
\end{center}
\vskip -0.2in
\end{figure*}

We conclude that our approach significantly outperforms
the rival algorithms in terms of the precision-recall
characteristic. The reason for this increase
in change-point detection performance is the high correlation
between the true change-points and the proposed detections,
as can be seen in Fig.~\ref{fig:trajectory_artificial}, right.
We note, however, that due to the complex nature of
both artificial datasets, many change-points are difficult to detect.

\begin{figure*}[h!t!]
\begin{center}
\begin{subfigure}{\textwidth}
\centerline{
\includegraphics[
    width=0.9\textwidth
]{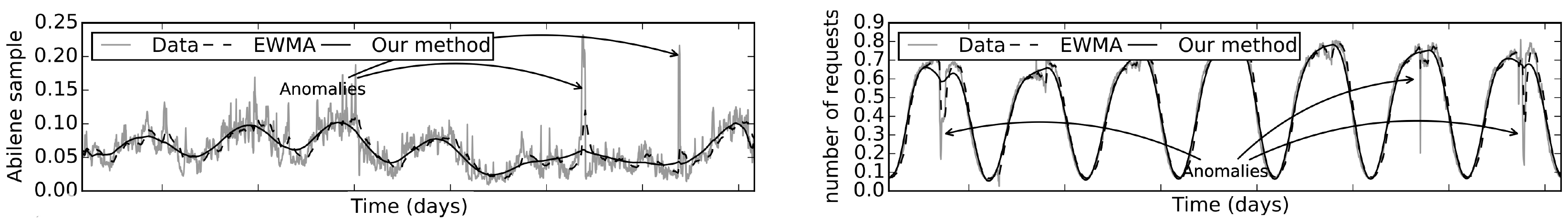}
}
\caption{Comparison with EWMA}
\label{fig:trajectory_real}
\end{subfigure}\\

\begin{subfigure}{\textwidth}
\centerline{
\includegraphics[
    width=0.9\textwidth
]{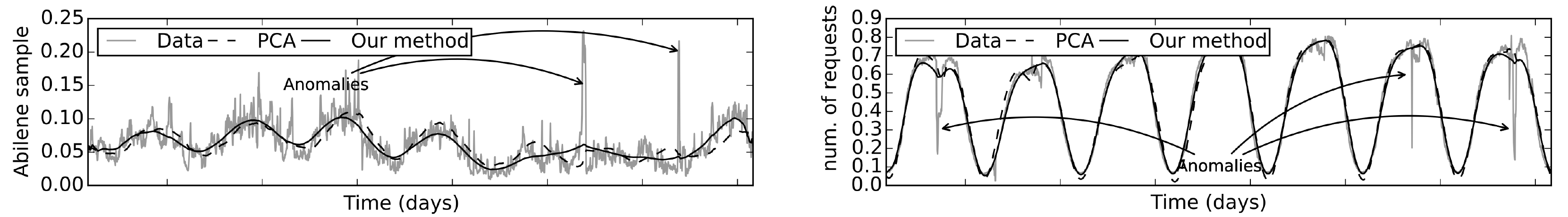}
}
\caption{Comparison with PCA}
\label{fig:trajectory_real_pca}
\end{subfigure}
\end{center}
\caption{Example data from the 
Abilene dataset (left) and the Yandex dataset (right), and the trend extraction results obtained using competing methods and our approach. Marked are the detected anomalies.}
\end{figure*}

\begin{figure*}[h!t!]
\begin{center}
\centerline{
  \includegraphics[
    width=0.55\textwidth]{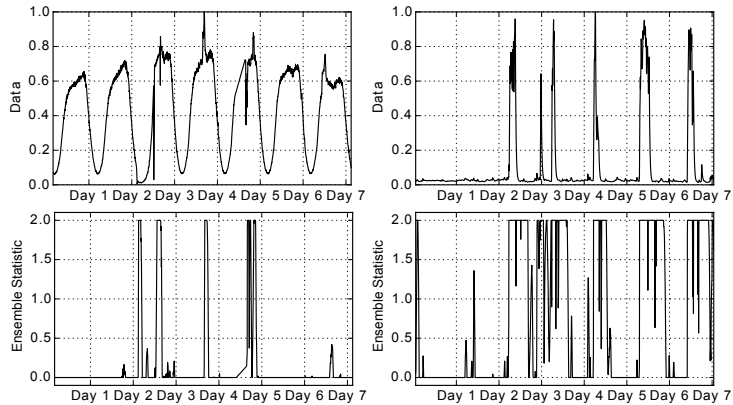}
}
\caption{Change-point detection results for two
different cases: Yandex, short-lived change, and Abilene, short-lived change
(left to right). From top to bottom are shown: the
source data $X_t$ and the logistic
regression-based ensemble statistic $a_t$.}
\label{fig:cusum_data_example}
\end{center}
\vskip -0.2in
\end{figure*}

Trend extraction results for the two real-world datasets are presented in
Fig.~\ref{fig:trajectory_real} for EWMA and our approach,
and in Fig.~\ref{fig:trajectory_real_pca} for PCA and our approach.
As can be seen from these figures, our filtering approach would
result in residuals which violate the change-point model
in~\eqref{eq:short_term_change_process} to a lesser extent;
the ensemble then further should improve detection performance
because it optimizes~\eqref{eq:ara_measures} on the residual data.
PCA-based approach performs generally comparable to our approach
(and even outperforms it in case of pretraining); however, PCA requires retraining
which is computationally very expensive when performed online
on a large number of time series. Our filtering approach is advantageous
in that it may be implemented online via a simple linear filter.
More results are in Fig.~\ref{fig:cusum_data_example},
where change-point detection results using the logistic regression-based
ensemble are presented for both Yandex and Abilene data.
We conclude that our approach is effective for both artificial
and real data and can readily be applied for anomaly
detection in a multitude of environments.

\section{Conclusion}
\label{conclusion}

We investigated change-point detection in the presence 
of quasi-seasonal trends and long-range dependent noise
with an application to fault detection in software-intensive systems.
We proposed an effective trend estimation algorithm
based on the theoretically optimal filter
and a practical change-point detection procedure
based on the ensemble of ``weak'' detectors.
An empirical study of the change-point detection
procedure shows that it significantly ourperforms
the standard EWMA and PCA-based algorithms when the conventional
assumptions about the change-point model are violated.

 \section{Acknowledgements}
 \label{acknowledgement}
The research, presented in Section \ref{applications} of this paper, was supported by the RFBR grants 16-01-00576 A and 16-29-09649 ofi\_m; the research, presented in other sections, was conducted in IITP RAS and supported solely by the Russian Science Foundation grant (project 14-50-00150).

\bibliographystyle{splncs}

\bibliography{references_v4}

\begin{thebibliography}{10}

\bibitem{iso42010}
ISO/IEC/IEEE:
\newblock Systems and software engineering -- architecture description.
\newblock ISO/IEC/IEEE 42010:2011(E) (Revision of ISO/IEC 42010:2007 and IEEE
  Std 1471-2000) (1 2011)  1 --46

\bibitem{northrop2006ultra}
Northrop, L., Feiler, P., Gabriel, R.P., Goodenough, J., Linger, R., Longstaff,
  T., Kazman, R., Klein, M., Schmidt, D., Sullivan, K.,  et~al.:
\newblock Ultra-large-scale systems: The software challenge of the future.
\newblock Technical report, DTIC Document (2006)

\bibitem{Yigitbasi2010}
Yigitbasi, N., Gallet, M., Kondo, D., Iosup, A., Epema, D.:
\newblock {Analysis and modeling of time-correlated failures in large-scale
  distributed systems}.
\newblock Proceedings - IEEE/ACM International Workshop on Grid Computing
  (2010)  65--72

\bibitem{Polunchenko2012}
Polunchenko, A.S., Tartakovsky, A.G.:
\newblock {State-of-the-Art in Sequential Change-Point Detection}.
\newblock Methodology and Computing in Applied Probability \textbf{14}(3)
  (2012)  649--684

\bibitem{Leland1994}
Leland, W.E., Taqqu, M.S., Willinger, W., Wilson, D.V.:
\newblock {On the self-similar nature of Ethernet traffic (extended version)}
  (1994)

\bibitem{Page1954}
Page, E.:
\newblock {Continuous inspection schemes}.
\newblock Biometrika \textbf{41}(1) (1954)  100--115

\bibitem{Shewhart1931}
Shewhart, W.A.:
\newblock {Economic control of quality of manufactured product} (1931)

\bibitem{Pham2014}
{Pham, Duc-Son and Venkatesh, Svetha and Lazarescu, Mihai and Budhaditya}, S.:
\newblock {Anomaly detection in large-scale data stream networks}.
\newblock Data Mining and Knowledge Discovery \textbf{28}(1) (2014)  145--189

\bibitem{Casas2010}
Casas, P., Vaton, S., Fillatre, L., Nikiforov, I.:
\newblock {Optimal volume anomaly detection and isolation in large-scale IP
  networks using coarse-grained measurements}.
\newblock Computer Networks \textbf{54}(11) (2010)  1750--1766

\bibitem{Lakhina2004}
Lakhina, A., Crovella, M., Diot, C.:
\newblock {Diagnosing network-wide traffic anomalies}.
\newblock ACM SIGCOMM Computer Communication Review \textbf{34}(4) (2004)  219

\bibitem{winters1960forecasting}
Winters, P.R.:
\newblock Forecasting sales by exponentially weighted moving averages.
\newblock Management Science \textbf{6}(3) (1960)  324--342

\bibitem{findley1998new}
Findley, D.F., Monsell, B.C., Bell, W.R., Otto, M.C., Chen, B.C.:
\newblock New capabilities and methods of the x-12-arima seasonal-adjustment
  program.
\newblock Journal of Business \& Economic Statistics \textbf{16}(2) (1998)
  127--152

\bibitem{hodrick1997postwar}
Hodrick, R.J., Prescott, E.C.:
\newblock Postwar us business cycles: an empirical investigation.
\newblock Journal of Money, credit, and Banking (1997)  1--16

\bibitem{artemov2015nonparametric}
Artemov, A.V., Burnaev, E.V., Lokot, A.S.:
\newblock Nonparametric decomposition of quasi-periodic time series for
  change-point detection.
\newblock In: Eighth International Conference on Machine Vision, International
  Society for Optics and Photonics (2015)  987520--987520

\bibitem{kolmogorov1940wiener}
Kolmogorov, A.N.:
\newblock The wiener spiral and some other interesting curves in hilbert space.
\newblock In: Dokl. Akad. Nauk SSSR. Volume~26. (1940)  115--118

\bibitem{Mandelbrot1968}
Mandelbrot, B.B., {Van Ness}, J.W.:
\newblock {Fractional Brownian Motions, Fractional Noises and Applications}
  (1968)

\bibitem{artemov2015optimal}
Artemov, A.V., Burnaev, E.V.:
\newblock Optimal estimation of a signal perturbed by a fractional brownian
  noise.
\newblock Theory Probab. Appl. \textbf{60}(1) (2016)  126--–134

\bibitem{kirichenko2011comparative}
Kirichenko, L., Radivilova, T., Deineko, Z.:
\newblock Comparative analysis for estimating of the hurst exponent for
  stationary and nonstationary time series.
\newblock Information Technologies \& Knowledge \textbf{5}(1) (2011)  371--388

\bibitem{hardstone2012detrended}
Hardstone, R., Poil, S.S., Schiavone, G., Jansen, R., Nikulin, V.V.,
  Mansvelder, H.D., Linkenkaer-Hansen, K.:
\newblock Detrended fluctuation analysis: a scale-free view on neuronal
  oscillations.
\newblock Scale-free Dynamics and Critical Phenomena in Cortical Activity
  (2012) ~75

\bibitem{artemov2015ensembles}
Artemov, A.V., Burnaev, E.V.:
\newblock Ensembles of detectors for online detection of transient changes.
\newblock In: Eighth International Conference on Machine Vision, International
  Society for Optics and Photonics (2015)  98751Z--98751Z

\bibitem{vautard1992singular}
Vautard, R., Yiou, P., Ghil, M.:
\newblock Singular-spectrum analysis: A toolkit for short, noisy chaotic
  signals.
\newblock Physica D: Nonlinear Phenomena \textbf{58}(1) (1992)  95--126

\end{thebibliography}

\end{document}